\documentclass[twocolumn,showpacs,preprintnumbers,amsmath,amssymb]{revtex4}

\usepackage{graphicx}% Include figure files
\usepackage{dcolumn}% Align table columns on decimal point
\usepackage{bm}% bold math

\begin{document}

\preprint{APS/123-QED}

\title{ Magnetocaloric effect in nano- and polycrystalline manganite $ La_{0.7}Ca_{0.3}MnO_3$}

\author{M. Pekala}
\email{pekala@chem.uw.edu.pl}
\affiliation{Department of Chemistry, University of Warsaw
Al. Zwirki i Wigury 101, PL-02-089 Warsaw, Poland}

\author{V. Drozd}
\email{pekala@chem.uw.edu.pl}
\affiliation{Department of Chemistry, University of Warsaw
Al. Zwirki i Wigury 101, PL-02-089 Warsaw, Poland, 
\\Department of Chemistry, Kiev National Taras Shevchenko University, 
60 Volodymyrska st, Kiev, 01033 Ukraine
\\Center for Study Matter at Extreme Conditions, Florida International University, 
Miami, FL 33199, USA}

\author{J.F. Fagnard }
\email{fagnard@montefiore.ulg.ac.be}
\affiliation{Montefiore Electricity Institute, B28,  Universit\'e de Li\`ege,
B-4000 Li`ege, Belgium}

\author{ Ph. Vanderbemden}
\email{ philippe.vanderbemden@ulg.ac.be }
\affiliation{Montefiore Electricity Institute, B28,  Universit\'e de Li\`ege,
B-4000 Li\`ege, Belgium}

\author{M. Ausloos}
\email{Marcel.Ausloos@ulg.ac.be}
\affiliation{GRAPES,  SUPRATECS, Universit\'e de Li\`ege, B5 Sart-Tilman, B-4000 Li\`ege, Euroland}

\date{09/07/2005}

\begin{abstract}
$ La_{0.7}Ca_{0.3}MnO_3$ samples were prepared in nano- and polycrystalline forms by sol-gel and solid state reaction methods, respectively, and structurally characterized by synchrotron X-ray diffraction. The magnetic properties determined by ac susceptibility and dc magnetization measurements are discussed. The magnetocaloric effect in this nanocrystalline manganite is spread over a broader temperature interval than in the polycrystalline case. The relative cooling power of the poly- and nanocrystalline manganites is used to evaluate a possible application for magnetic cooling below room temperature.
\end{abstract}


\maketitle

% main text
\section{Introduction}
Mixed valence manganites with perovskite structure have been studied for more than 50 years due to the interesting interplay between electronic and magnetic features. Beyond the colossal magnetoresistance discovered in 1994 some manganite systems exhibit also a relatively strong magnetocaloric effect at the ferro Ð paramagnetic transition temperature [1-3]. The magnitude of the magnetocaloric effect depends on the chemical composition, the $Mn^{3+}$ $Mn^{4+ }$ ratio and the microstructure. As the magnetocaloric effect may be applied for the contemporary cooling systems, magnetic and structural investigations of various manganites are needed.
	This paper presents results of new investigations on new magnetocaloric materials which might have properties applicable at room temperature. The report extends investigations [4-7] of the magnetocaloric effect in manganites and compares the effect of microstructure in nano- and polycrystalline $ La_{0.7}Ca_{0.3}MnO_3$  systems with that of related materials.
. 
 
\section{Sample synthesis and characterization}

$ La_{0.7}Ca_{0.3}MnO_3$  samples were prepared in nano- and polycrystalline forms by sol-gel and solid state reaction methods, respectively. Sol-gel method [8] started from $La_2O_3$, $CaCO_3$ and $MnO$ reagents which were dissolved in diluted nitric acid at continuous stirring and moderated heating. Prior to using the lanthanum oxide was calcined at 950oC for 24h to remove hydroxide and carbonate impurities. Gelation agent, monohydrate of citric acid, was added together with ethylene glycol. The molar ratios metals:citric acid:ethylene glycol were 1:10:10. The obtained solution was evaporated on a hot plate till a homogeneous gel-like product was formed, which was later decomposed at 300 C in air. Nanocrystalline $ La_{0.7}Ca_{0.3}MnO_3$  powder was produced by calcination of the sol-gel precursor at 600 C for 12 h in air atmosphere.

A polycrystalline $ La_{0.7}Ca_{0.3}MnO_3$  sample was prepared by the solid state reaction method using the same starting reagents. The mixture of reagents was calcined at 900oC in air for 24h, grinded and sintered at 1000 C and 1100 C for 24h at each temperature with intermediate grinding at room temperature.

`Synchrotron XRD experiments were carried out in the BL01C beam line (National Synchrotron Radiation Research Center (NSRRC), Taiwan) by using a ACCM Si(111) monochromator at a synchrotron wavelength of 0.0516606 nm. The integrated 1D patterns were processed by FIT2D program. The diffraction angles were calibrated according to the NBS 640b standard sample Si powder. Structural refinements were made with X-ray data  by using GSAS program [9].

Samples are single phase and orthorhombic at room temperature. The unit cell parameters listed in Table 1 show that the unit cell is slightly larger for the polycrystalline than for the nanocrystalline sample. A similar tendency to reduction of the unit cell volume of a nanocrystalline $ La_{0.7}Ca_{0.3}MnO_3$  manganite was also reported by Rozenberg et al. [10]. The oxygen deficiency equals to 0.01 according to iodometric titration.

For the grain size determination the Lanthanum hexaboride powder of particle size $\sim$2 $\mu$$m$ was used to determine the instrumental contribution to diffraction peaks broadening. XRD patterns of manganites and $LaB_6$ were recorded at the same conditions. Profile fitting of XRD patterns was performed using pseudo-Voigt function in GSAS. Profile parameters of manganites were set to those ones of $LaB_6$ except for Lorentzian Scherrer broadening (X) term which was refined for both poly- and nanocrystalline samples (Fig. 1). The grain size (in $0.1 nm$) can be calculated from Scherrer equation.

%??????
%\begin{equation} 
% P=\frac{18000 B}{\pi \chi'} 
%\end{equation}
%???????????

%where $B$ is Scherrer constant (0.9); $===$ is X-ray radiation wavelength; $===$ Ð Lorentzian Scherrer broadening term in GSAS profile function after subtration of instrumental contribution.

\begin{table}

\caption{Structural parameters of  $ La_{0.7}Ca_{0.3}MnO_3$  samples based on Rietveld refinement in the orthorhombic space group $Pnma$ (No. 62) of synchrotron radiation ($\lambda$ = 0.0516606 nm) XRD patterns measured at 300K.}
\smallskip
\begin{footnotesize}
\begin{center}
\begin{tabular}{c c c c c c c c c c}

\\\hline 
\\  {\bf cell } & {\bf Nanocrystalline} &	{\bf Polycrystalline}\\
\\  {\bf parameter} &{\bf $ La_{0.7}Ca_{0.3}MnO_3$  }&	{\bf $ La_{0.7}Ca_{0.3}MnO_3$}  \\
\\$a$, nm &	0.5435(6) &	0.5473(3)\\
\\$b$, nm&0.7691(8) &	0.7739(4)\\
\\$c$, nm&	0.5436(6)&	5.479(3)\\
\\$V$, nm$^3$&	0.2272(5)&	0.2320(2)\\
\\\hline

\end{tabular}
\end{center}
\end{footnotesize}
\end{table}

The mean grain sizes of the nanocrystalline and polycrystalline samples are around 20 nm and 100 nm, respectively, as determined from XRD measurements. According to scanning electron microscopy observations the particle mean size of polycrystalline  $ La_{0.7}Ca_{0.3}MnO_3$  sample was below 0.5 $\mu$$m$, when determined from two dimensional micrographs. The electron microscopy is known to be more sensitive to the relatively larger objects. Therefore, it was not possible to achieve enough contrast in micrographs of the nanocrystalline manganite.

\section{Measurements }
The temperature variation of the magnetic AC susceptibility and DC magnetization were searched for magnetic characteristics by means of the PPMS system. The temperature dependence of the AC magnetic susceptibility was measured at 1 mT amplitude.

 The temperature variation of the field-cooled (FC) and zero-field-cooled (ZFC) magnetization were registered at magnetic fields of 0.01 and 0.03 T. The DC magnetization isotherms were recorded in magnetic fields up to 2 T.

 \begin{figure}
\hspace{1.8cm}
\includegraphics[width=3.5in]{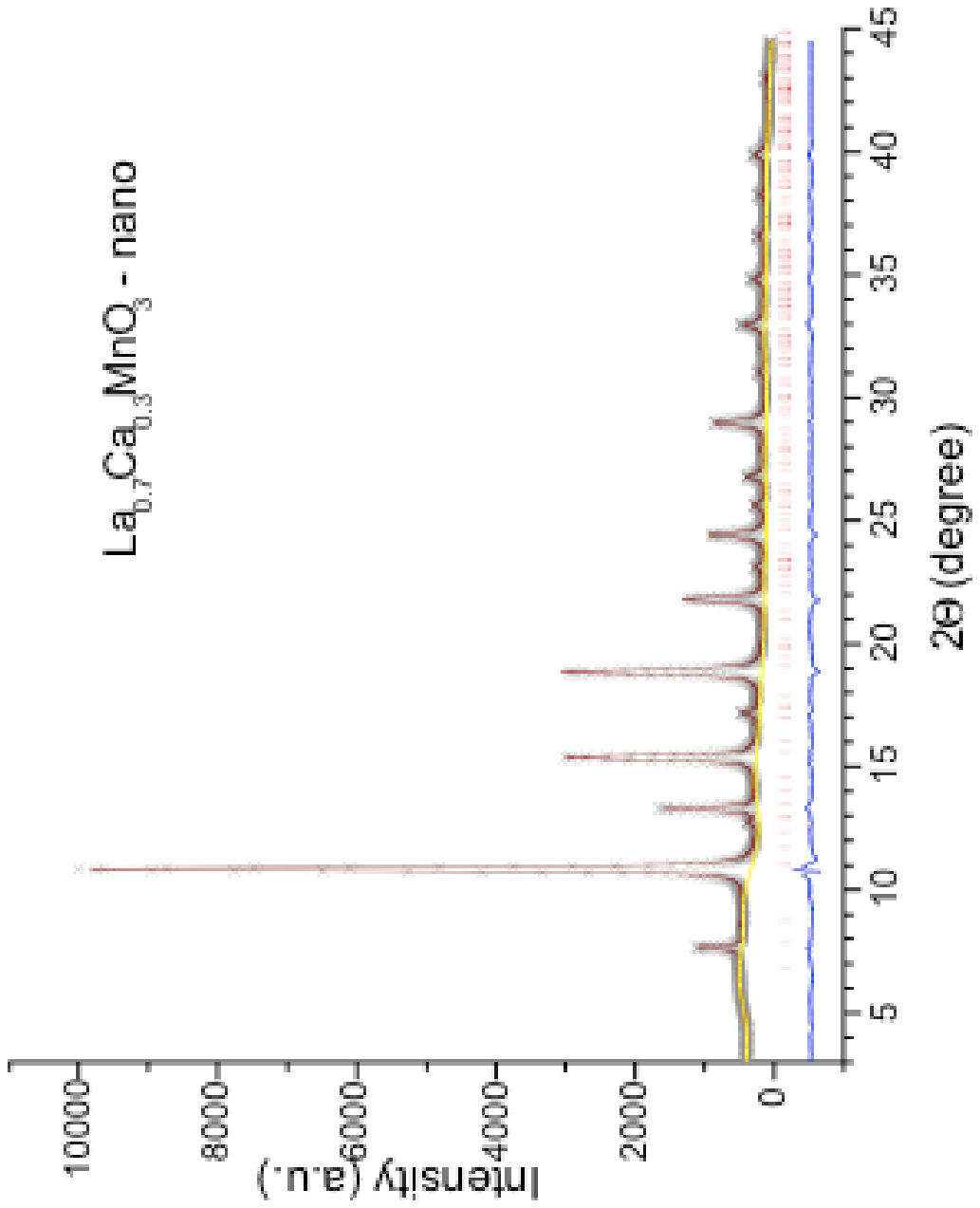}
\includegraphics[width=3.5in]{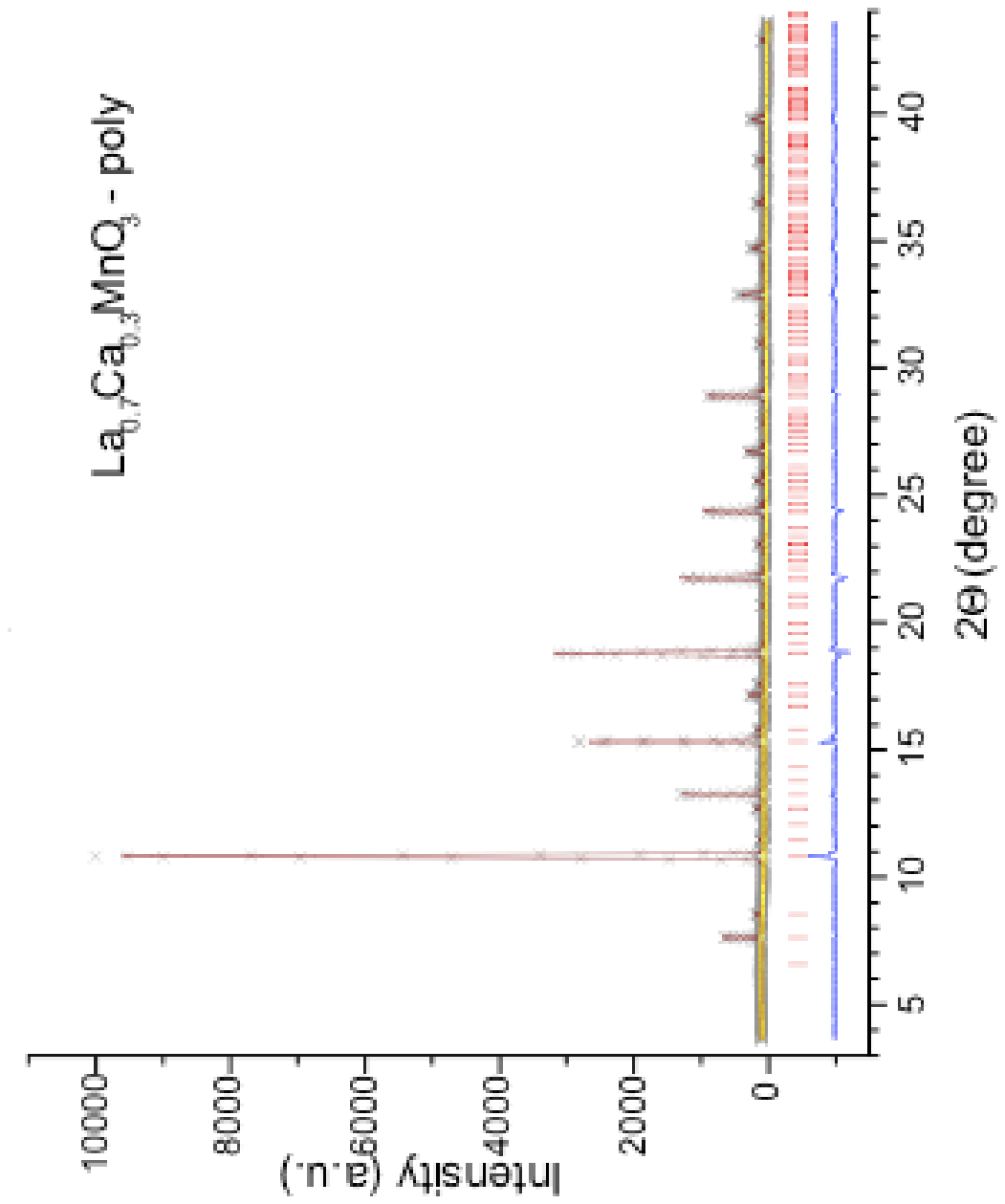}
\caption{\label{fig1}Rietveld refinement plot for nanocrystalline (a), polycrystalline (b) powders of$ La_{0.7}Ca_{0.3}MnO_3$ (observed, +; calculated, - (solid line); reflections, -(vertical lines); difference bottom)
  }
\end{figure}

\section{Magnetic results}

The temperature dependencies of the in-phase component  of the AC magnetic susceptibility $\chi'$ (Fig. 2) reveal that both poly- and nanocrystalline manganites remain ferromagnetic up to the Curie temperature of 266.3 and 239.3 K, respectively, as defined by the largest slope of $\chi'(T)$. The shape of the $\chi'(T)$ curves confirms that the samples are single phase. The magnitude of the susceptibility is almost 3 times smaller for the nano- than poly-crystalline manganite. The susceptibility maximum observed around 160 K for the polycrystalline sample is found at 100 K in the nanocrystalline case. The high temperature drop of the magnetic susceptibility of the polycrystalline manganite occurs in a relatively narrower temperature interval (30 K) than for the nanocrystalline one (50 K). The low temperature susceptibility decay is approximately five times more abrupt in the nanocrystalline sample. Absolute values of the out-of-phase susceptibilities reveal that the magnetic energy dissipation is approximately 20 percent higher in the nanocrystalline manganite as compared to the polycrystalline one. Such a difference reflects a higher structural and magnetic disorder on grain boundaries. The maxima in out-of-phase components $\chi''(T)$ of the AC magnetic susceptibility are located around 210 K independent of the microcrystalline form. The abrupt decays of the magnetic susceptibility hint towards a possibly strong magnetocaloric effect in the transition temperature range. The field-cooled (FC) and zero-field-cooled (ZFC) magnetization measurements at magnetic fields of 0.01 and 0.03 T (not plotted) confirm values of Curie temperatures determined from magnetic susceptibility.

 \begin{figure}
\hspace{1.8cm}
\includegraphics[width=3.5in]{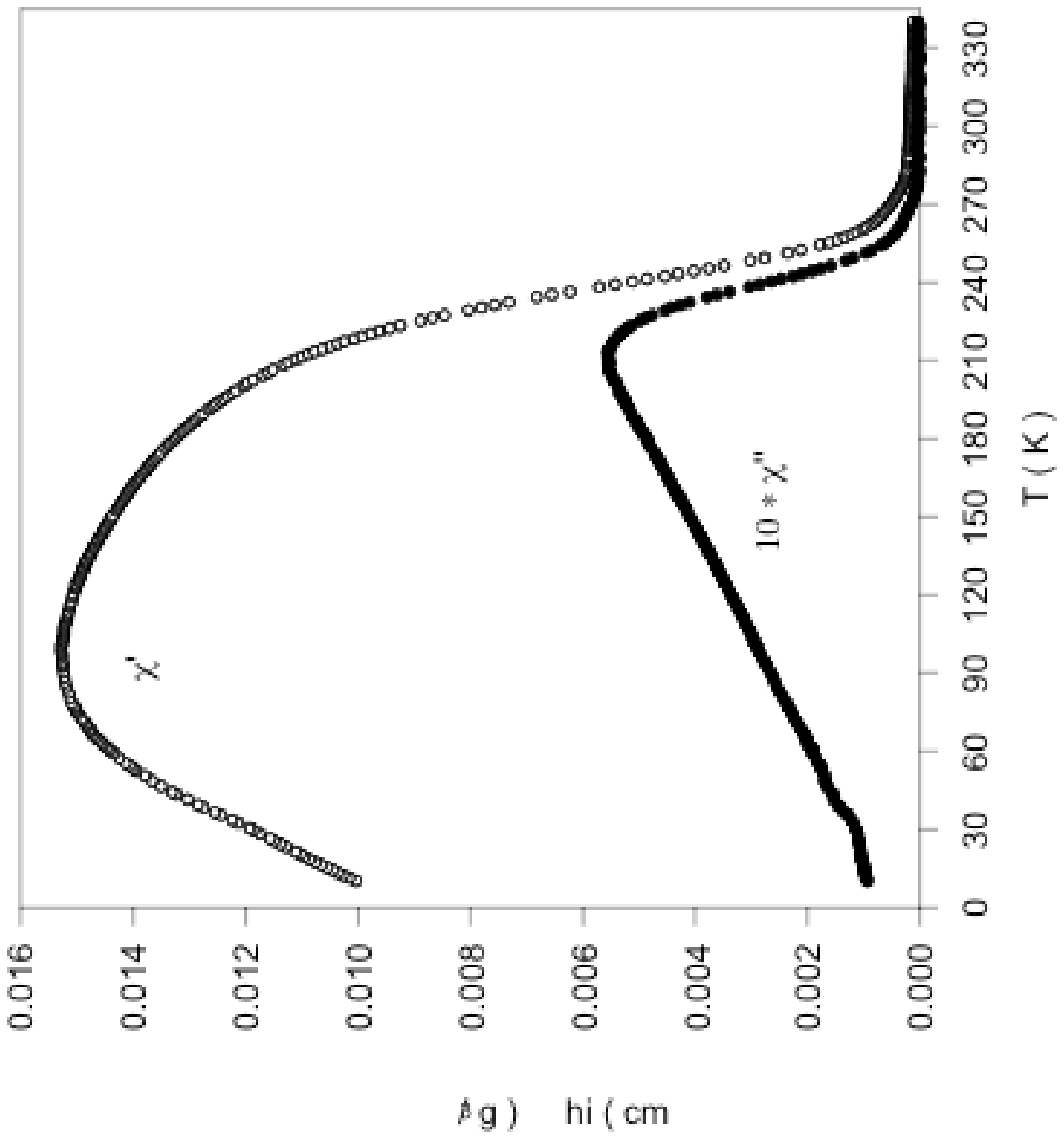}
\includegraphics[width=3.5in]{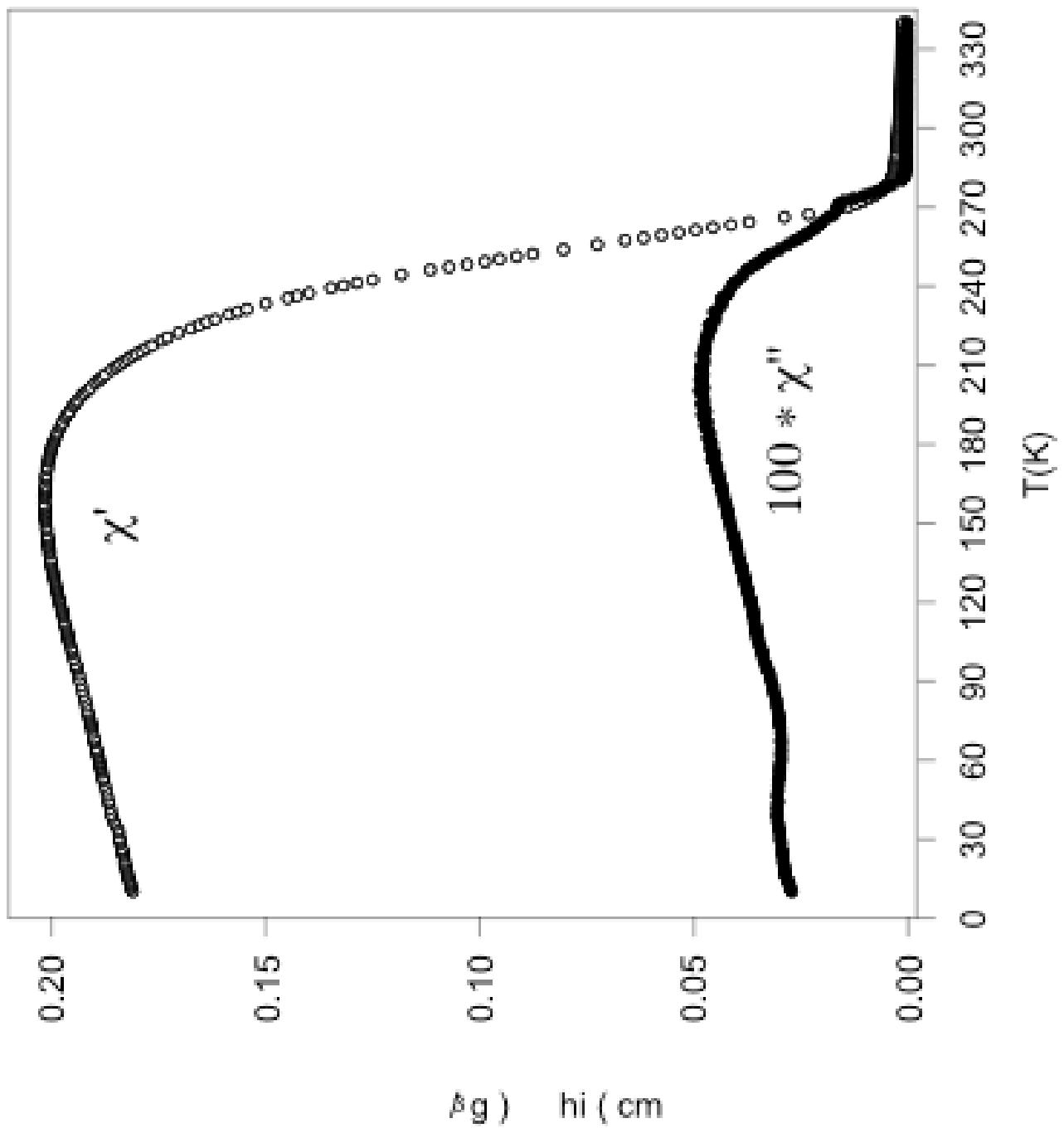}
\caption{\label{fig2} Temperature variation of AC magnetic susceptibility for nano- (a) and polycrystalline (b) manganites$ La_{0.7}Ca_{0.3}MnO_3$ }
\end{figure}
  
  In order to study an influence of microstructure on the magnetic entropy change, the temperature and field dependence of magnetization was measured. The DC magnetization isotherms only approach the magnetic saturation state (Fig. 3). However, a 2 T magnetic field does not saturate either the poly- or nano-crystalline manganites at 192 and 150 K, respectively. For the polycrystalline manganite the experimental values of the magnetic moment per formula unit at 2 T and 192 K equal to 3.0 $\mu$$B$ , is lower than the theoretical Òspin-onlyÓ magnetic moment of 3.7 $\mu$$B$, calculated assuming that magnetic moments per $Mn^{3+}$ and $Mn^{4+}$ ions equal to 4 $\mu$$B$ and 3 $\mu$$B$, respectively. The magnetic moment is even lower for the nanocrystalline manganite, achieving only 1.1 $\mu$$B$, at 150 K. The temperature variations of the spontaneous magnetization and the inverse DC magnetic susceptibility derived from magnetization isotherms are plotted in Fig. 4. They indicate a Curie temperature value in agreement with values derived from AC susceptibility results. For the nanocrystalline manganite the nonlinear shape of the inverse susceptibility vs. temperature dependence below 240 K suggests additional investigations.

 \begin{figure}
\hspace{1.8cm}
\includegraphics[width=3.5in]{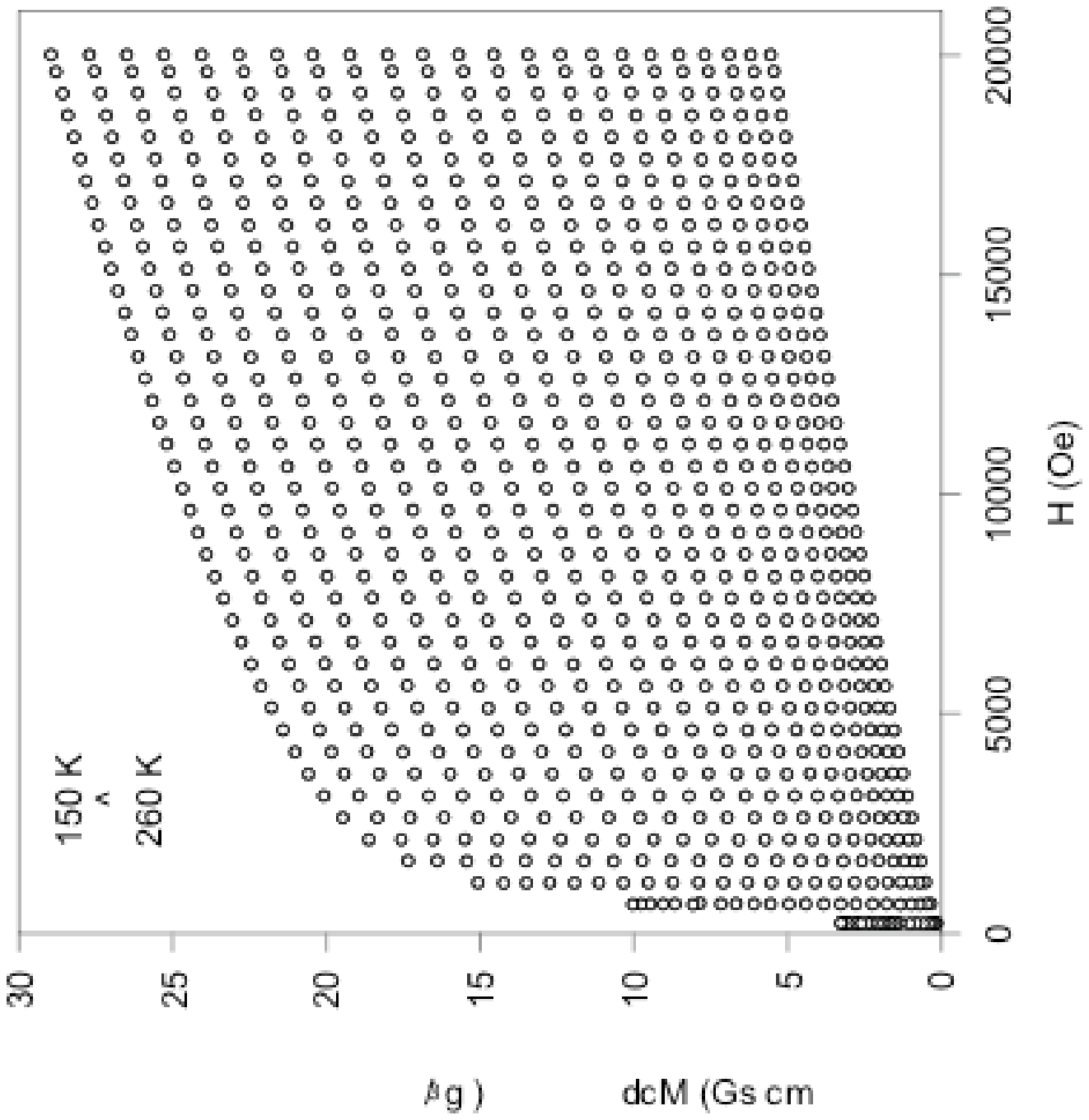}
\includegraphics[width=3.5in]{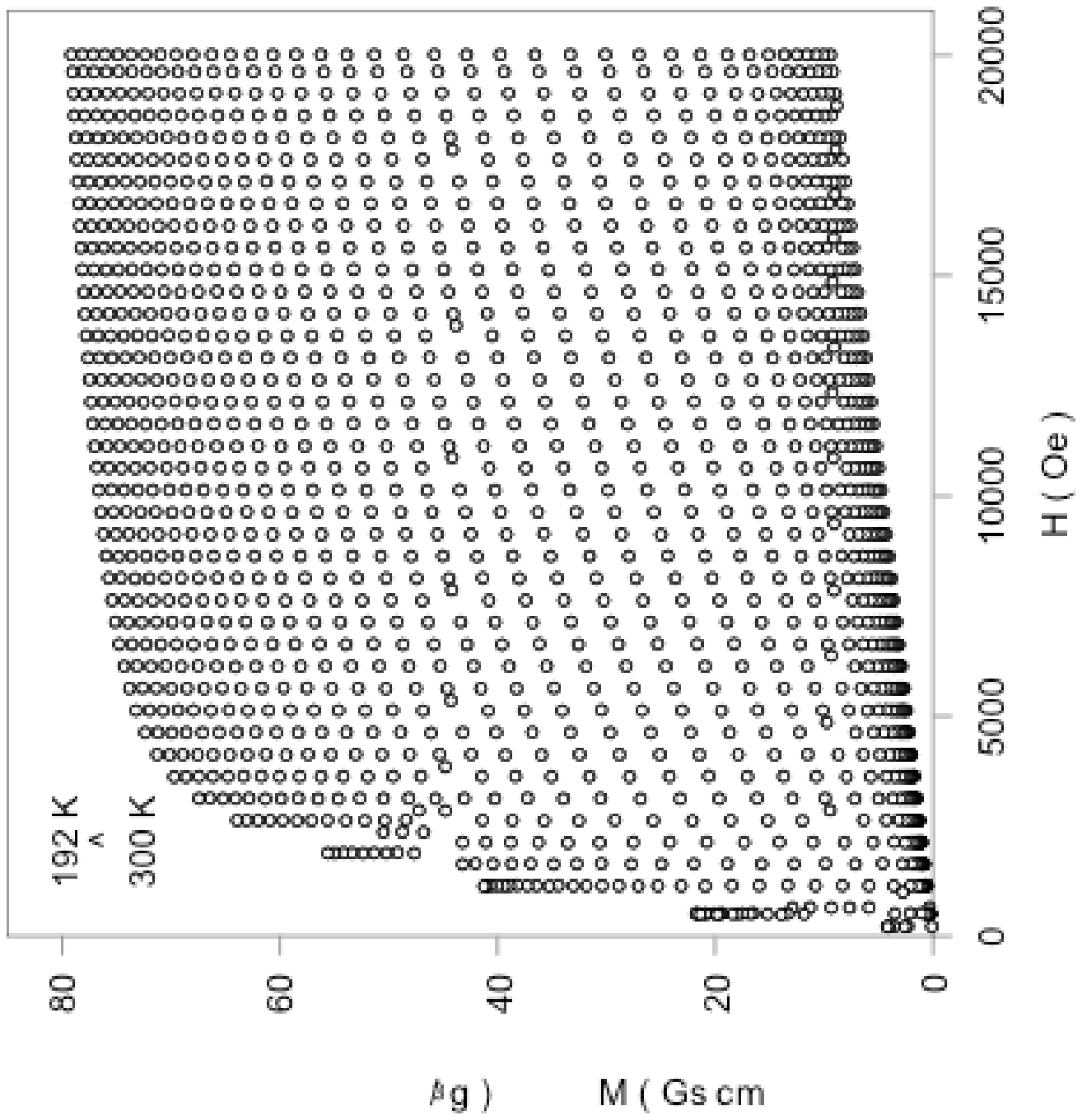}
\caption{\label{fig3} The DC magnetization isotherms for nano- (a) and polycrystalline (b) manganites $ La_{0.7}Ca_{0.3}MnO_3$  at various temperatures. }
\end{figure}

 \begin{figure}
\hspace{1.8cm}
\includegraphics[width=3.5in]{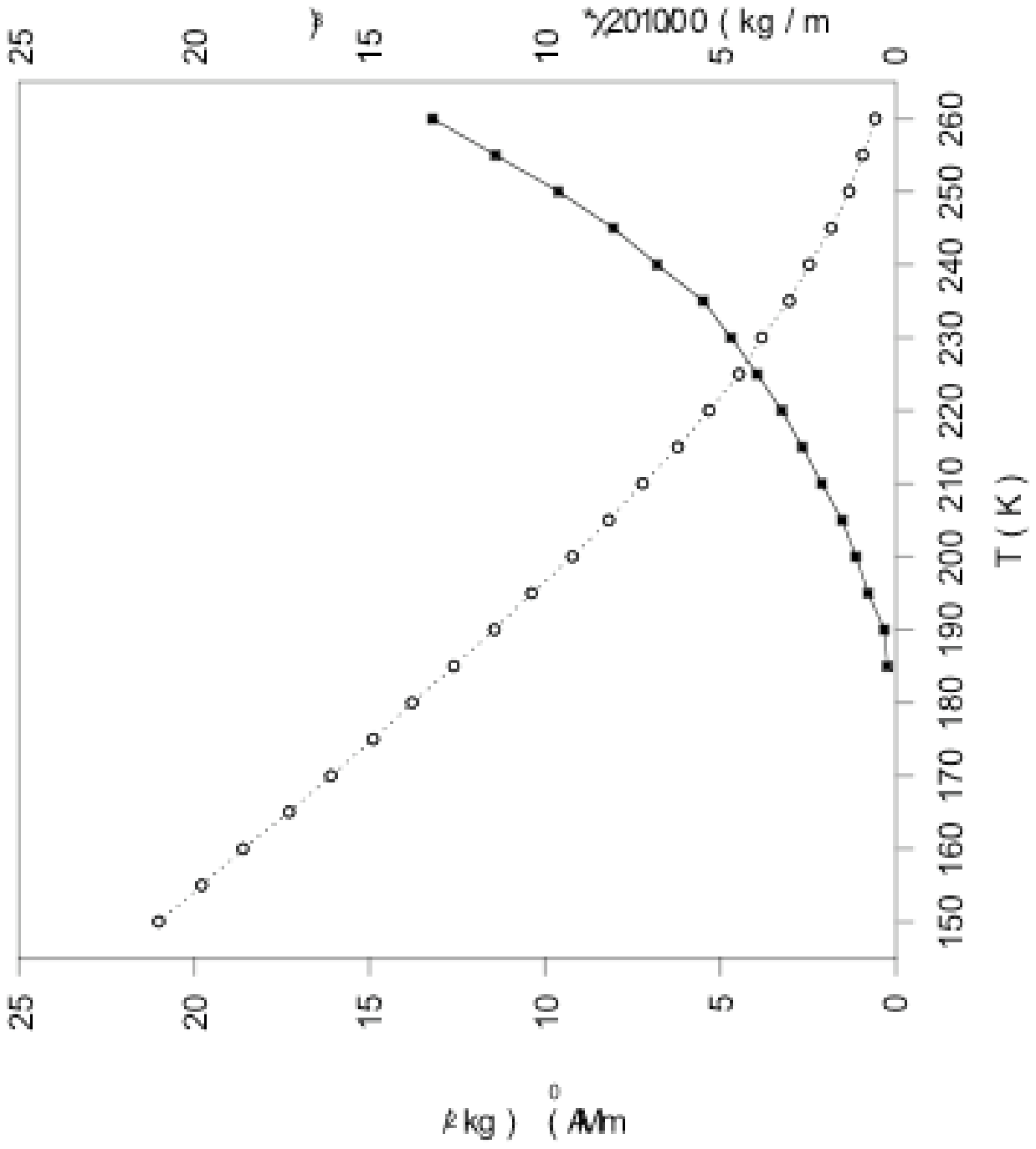}
\includegraphics[width=3.5in]{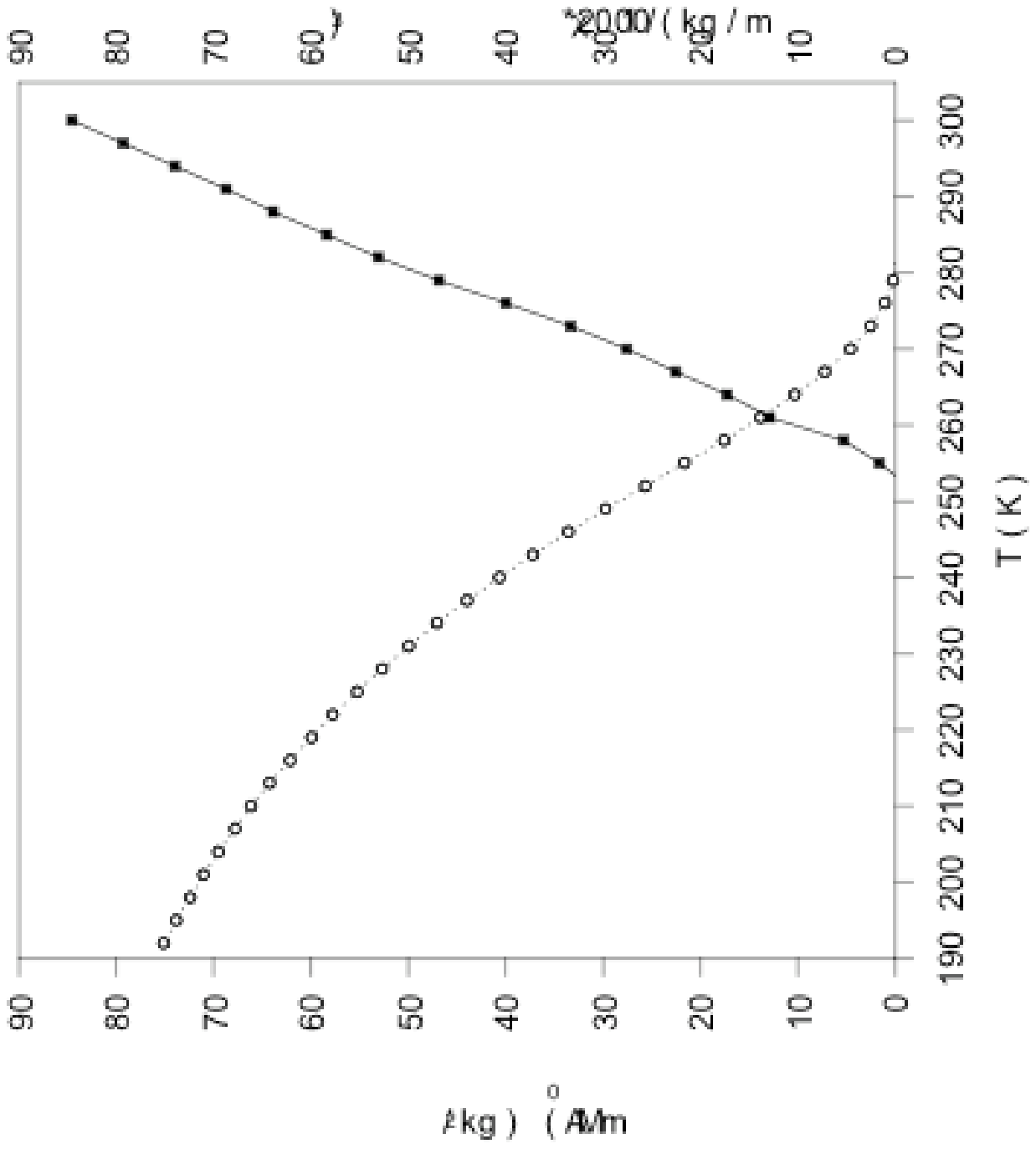}
\vspace{-0.5cm}
\caption{\label{fig4} Temperature variation of the spontaneous magnetization (open symbol) and inverse magnetic susceptibility (solid symbol) for nano- (a)  and polycrystalline (b) manganite $La_{0.7}Ca_{0.3}MnO_3$  }
\end{figure}

 \begin{figure}
\hspace{1.8cm}
\includegraphics[width=3.5in]{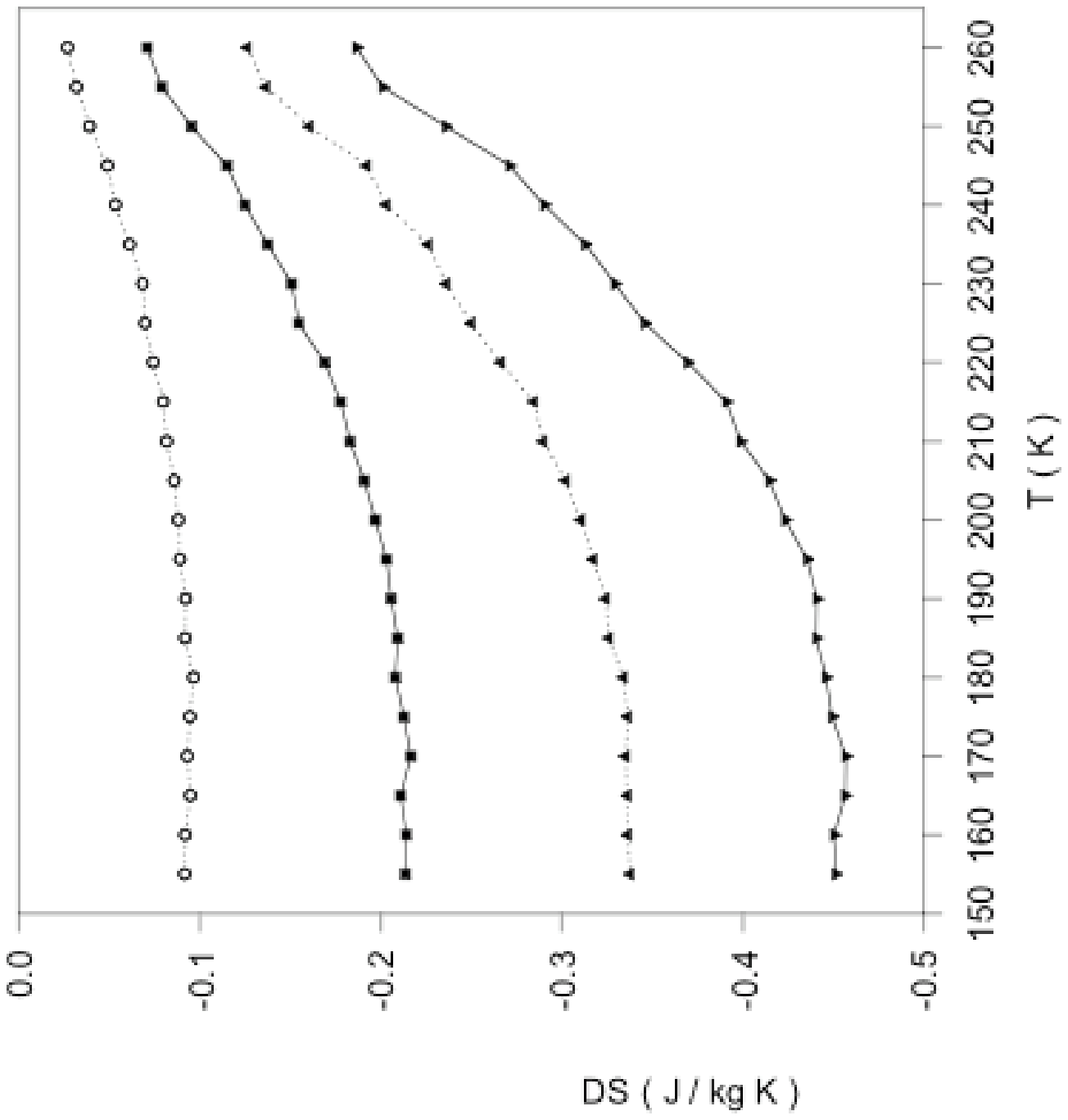}
\includegraphics[width=3.5in]{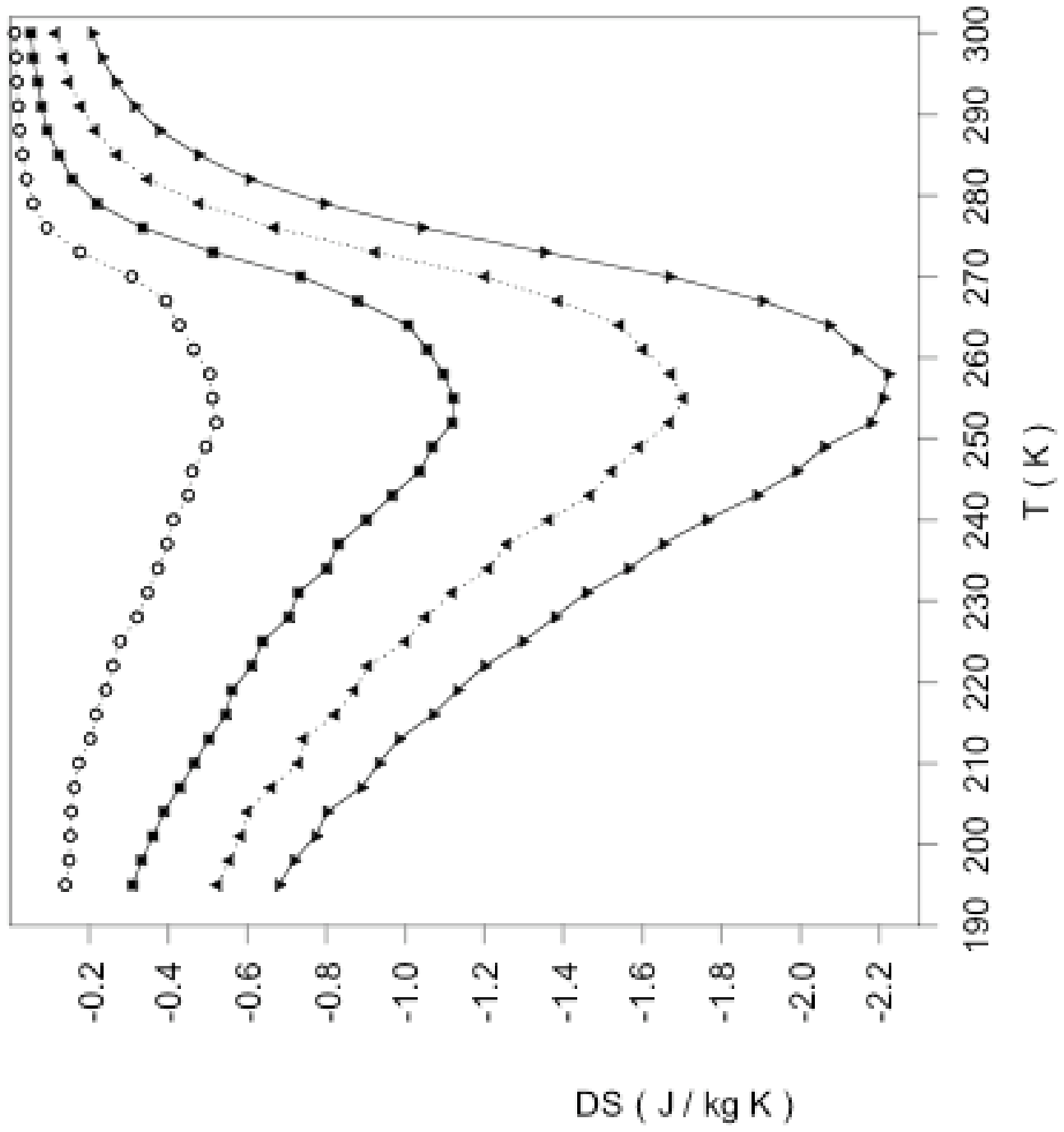}
\caption{\label{fig5} Magnetic entropy change around the Curie temperature for nano- (a) and polycrystalline (b) manganites $ La_{0.7}Ca_{0.3}MnO_3$  at magnetic fields from 0.5 T (top) to 2 T (bottom).}
\end{figure}

  \section{Magnetocaloric effect}
  
  The temperature variation of the magnetic entropy change,($DS$), derived from magnetization isotherms [1-3] are plotted in Fig. 5. Remarkable differences can be observed between the poly- and nanocrystalline manganites. The minimum value of $DS $ for the polycrystalline manganite is located at 252 K at 0.5 T magnetic field (Fig. 5b). The magnitude of this minimum increases approximately proportionally to the magnetic field and reaches -2.2 J/kg K at 2 T. This is accompanied by a broadening of the $DS$ peak. The latter shifts towards a higher temperature (258 K) when the magnetic field increases from 0.5 to 2 T. The $DS$ vs. temperature curve is slightly asymmetric with a faster variation on the high temperature side. This was also observed in other polycrystalline manganites and ascribed to a grain boundary structural effect [11].
  
 	Another intriguing feature is that the $DS$ peak appears somewhat above the Curie temperature of the polycrystalline manganite. This feature seems to be related to a specific sample preparation procedure since the magnetic susceptibility and magnetization measurements were made on the same sample and using the same thermometer.
	
	For the nanocrystalline manganite the magnetocaloric effect is rather low at 260 K reaching 0.2 J/kg K at 2 T (Fig. 5a). The magnitude of the magnetocaloric effect increases under decreasing temperature and becomes roughly constant below 190 K, for all magnetic fields applied. One may notice that below 190 K, the magnitude of the magnetic entropy change is five times smaller than the $DS$ peak of the polycrystalline manganite.

  	Due to scarce reports the magnetocaloric effect in the polycrystalline manganite may be compared only to results of Phan et al. [11-13]. One may notice that the magnitude of the minimum $DS$ value for the polycrystalline manganite studied is about 30 percent lower than that reported for the single and polycrystalline samples [11,12]. Moreover, at 1 T magnetic field the $DS$ peak is located about 12 K below the one reported by Phan et al. [11] for the polycrystalline manganite. On the other hand, one may remind that the $DS$ peak for the $La_{0.7}Ca_{0.3}MnO_3$  single crystal appears at considerably lower Curie temperature of 227 K [11]. As it is usually observed for manganites, the highest magnetocaloric effect is found in single crystals of corresponding chemical composition [14].
	
	The cooling efficiency of magnetic refrigerant is estimated by the so called relative cooling power $RCP(S)$ corresponding to the amount of heat transferred between the cold and hot sinks in the ideal refrigeration cycle [1-3] is written as follows
\begin{equation} 
RCP(S) = DS_{MAX}  DT_{FWHM}	  
\end{equation}

where $DS_{MAX} $  is the $DS$ minimum and $DT_{FWHM}	 $ is the full width at half the minimum. Since the relatively broad $DS$ minimum extends over about 35 K, the relative cooling power $RCP(S)$ of the polycrystalline manganite $La_{0.7}Ca_{0.3}MnO_3$  is about 55 J/kg, which is 10 percent  larger than a value reported by Tian et el. [12]. Then assuming a density of 5.95 g/cm$^3$ one arrives to a $RCP(S)$ per cubic centimeter of material equal to 327 mJ/cm$^3$ at 2 T. This value is more than one fifth of the magnetocaloric standard of metallic $Gd$ at 2 T magnetic field [15]. Therefore it is shown that the polycrystalline  $ La_{0.7}Ca_{0.3}MnO_3$ manganite may be potentially considered as an interesting refrigerant not far below room temperature.

The $RCP(S)$ parameter of the nanocrystalline manganite  $ La_{0.7}Ca_{0.3}MnO_3$ cannot be precisely estimated using an approach described just above. The entropy change $DS$ in the nanocrystalline manganite is small indeed. However, the magnetocaloric effect for this sample is spread over a remarkably broader temperature interval. For example, at 2 T magnetic field, the $D$S is equal to 0.32 J/kg K corresponding to the level of full width at half minimum, but spreads over more than 80 K (between 150 and 230 K, see Fig. 5). That points to the possibility that a reduction in $RCP(S)$ due to the magnitude of $DS$, may be compensated by a much wider temperature interval.

According to previous reports the ferro- to paramagnetic phase transition in  $La_{0.7}Ca_{0.3}MnO_3$manganite evolves from the first to the second order when the gran size is suppressed towards nanometers [11,16]. The diminishing abruptness of phase transition is revealed by the increasing temperature interval where the magnetization decays around the Curie temperature. Therefore the narrow ( about 10 to 15 K ) magnetocaloric effect is reported for single crystals [11,12,14] and even for manganites with 500 nm grains [16]. The present results correlate with this tendency as indicated by temperature width ( 35 K ) of magnetocaloric effect for a polycrystalline manganite with grain size about 100 nm.

The large width of magnetocaloric effect in the nanocrystalline manganite is obviously related to a high degree of structural disorder which involves strong local stress in $MnO_6$ octahedra modifying the $Mn-O-Mn$ angles responsible for the lattice, magnetic and electronic properties. A thickness of the structurally disordered layer on grain surface is assumed to be roughly similar both in poly- and nanocrystalline manganites. Thus, in the nanocrystalline manganite with mean size about 20 nm, the remarkable volume fraction located in surface layer plays a deciding role in a broadening of magnetocaloric effect. 

Generally, the magnetocaloric effect in various families of materials is known to occur in a relatively narrow temperature range [1-3]. This in turn causes that special approaches, as sandwich systems composed of materials with a sequence of transition temperatures, are envisaged in order to enhance the temperature difference between the heat source and heat sink. In the case of this nanocrystalline  $La_{0.7}Ca_{0.3}MnO_3$ manganite, the relatively broad temperature range, where the magnetocaloric effect is meaningful, allows to overcome this limitation, especially when the magnitude of $DS$ may be increased by microstructual and technological factors.

{\bf Acknowledgements}
Work supported in parts by Ministry of Science and Higher Education (PL - grant WAL/286/2006), FNRS (BE), Kasa Mianowskiego (PL) and CGRI (BE).

 \newpage
 
 Figure captions

Fig. 1. Rietveld refinement plot for nanocrystalline (a), polycrystalline (b) powders of$ La_{0.7}Ca_{0.3}MnO_3$ (observed, +; calculated, - (solid line); reflections, -(vertical lines); difference bottom). \\

Fig. 2. Temperature variation of AC magnetic susceptibility for nano- (a) and polycrystalline (b) manganites$ La_{0.7}Ca_{0.3}MnO_3$   \\

Fig. 3. The DC magnetization isotherms for nano- (a) and polycrystalline (b) manganites $ La_{0.7}Ca_{0.3}MnO_3$  at various temperatures. \\

Fig. 4. Temperature variation of the spontaneous magnetization (open symbol) and inverse magnetic susceptibility (solid symbol) for nano- (a)  and polycrystalline (b) manganite $ La_{0.7}Ca_{0.3}MnO_3$  \\

Fig. 5. Magnetic entropy change around the Curie temperature for nano- (a) and polycrystalline (b) manganites $ La_{0.7}Ca_{0.3}MnO_3$  at magnetic fields from 0.5 T (top) to 2 T (bottom).

\end{document}